\documentclass[12pt]{article}           
\def\preprint{}       
\def\finished{}
\def\archive {hep-th/0109164}           

\def\title{Non-perturbative Gauge Groups and Local Mirror Symmetry}

\long\def\abstract{
We analyze D-brane states and their central charges on the resolution
of $\IC^2/\IZ_n$ by using local mirror symmetry. 
There is a point in the moduli space where all $n(n-1)/2$
branches of the principal component of the discriminant locus coincide.
We argue that this is the point where compactifications of Type IIA
theory on a K3 manifold containing such a local geometry acquire a
non-perturbative gauge symmetry of the type $A_{n-1}$.
This analysis, which involves an explicit solution of the GKZ system
of the local geometry, explains how the quantum geometry exhibits all
positive roots of $A_{n-1}$ and not just the simple roots that
manifest themselves as the exceptional curves of the classical geometry.
We also make some remarks related to McKay correspondence.

 }

\usepackage{amssymb,latexsym}              \catcode`\"=\active \let"=\"

\def\ifundefined#1{\expandafter\ifx\csname#1\endcsname\relax}
\def\bye{\end{document}}   
\long\def\new#1\endnew{{\bf #1}}
\long\def\del#1\enddel{} 

\def\HS#1 {\hspace*{#1pt}} \def\VS#1 {\vspace*{#1pt}}
\def\BC{\begin{center}}    
\def\EC{\end{center}}

\let\0=\over      \let\1=\vec      \def\2{{1\over2}}    \let\3=\ss
\let\4=\underline \let\5=\overline \let\6=\partial      \def\7#1{{#1}\llap{/}}
\def\8#1{{\textstyle{#1}}}         \def\9#1{{\ifmmode{\pmb{#1}}\else\bf#1\fi}}

          \def\({\left(}       \def\){\right)}

\def\eeql#1 {\label{#1}\eeq}      \let\nn=\nonumber  
\def\beq{\begin{equation}}      \def\eeq{\end{equation}}        
\def\bea{\begin{eqnarray}}      \def\eea{\end{eqnarray}}

\let\and=\wedge                   
\let\then=\Rightarrow      
                
\let\bra=\langle        \let\ket=\rangle        \def\<#1\>{\bra #1 \ket}
\let\ni=\noindent

\def\rel#1 #2{\buildrel #1 \over {#2}}

\let\a=\alpha   \let\b=\beta       \let\d=\delta   
           
   \let\l=\lambda        
            \let\p=\pi

\let\P=\Pi

  \def\co{{\cal O}}

 \def\IC{{\mathbb C}} \def\IP{{\mathbb P}} 
   
\def\IZ{{\mathbb Z}}

\def\plb#1 #2 {Phys. Lett. {\bf B#1} #2 }
\def\phr#1 #2 {Phys. Rep. {\bf  #1} #2 }        
\def\npb#1 #2 {Nucl. Phys. {\bf B#1} #2 }
\def\aph#1 #2 {Ann. Phys. {\bf #1} #2 }         
\def\jmp#1 #2 {J. Math. Phys. {\bf #1} #2 }
\def\jgp#1 #2 {J. Geom. Phys. {\bf #1} #2 }
\def\prd#1 #2 {Phys. Rev. {\bf D#1} #2 }
\def\prl#1 #2 {Phys. Rev. Lett. {\bf #1} #2 }
\def\rmp#1 #2 {Rev. Mod. Phys.  {\bf #1} #2 }
\def\zpc#1 {Z. Phys. {\bf #1C} }
\def\cmp#1 #2 {Commun. Math. Phys. {\bf #1} #2 }
\def\cqg#1 #2 {Class.Quant.Grav. {\bf #1} #2 }
\def\mpl#1 {Mod. Phys. Lett. {\bf A#1} }
\def\cpc#1 {Computer Phys. Commun. {\bf #1} }   
\def\ijmp#1 {Int. J. Mod. Phys. {\bf A#1} }
\def\ijmpC#1 {Int. J. Mod. Phys. {\bf C#1} }
\def\atmp#1 {Adv. Theor. Math. Phys. {\bf #1} }

\def\BP{\begin{picture}} \def\EP{\end{picture}}         
\newcounter{TRefNX} \let\OLDcite=\cite  \makeatletter
\def\makeTRefs#1{\@for  \NewTRef:=#1\do{\global\makeTRef{\NewTRef}}}
\def\makeTRef#1{\ifundefined{TRef#1}\stepcounter{TRefNX}%
\expandafter\xdef\csname TRef#1\endcsname{\theTRefNX}\fi}\makeatother
\def\NEWcite#1{\makeTRefs{#1}\OLDcite{#1}}  

\ifundefined{draftmode} {} 
\else        
   \let\cite=\NEWcite
   \newcount\HOUR  \HOUR=\the\time \divide\HOUR by 60   \multiply \HOUR by 60 
   \newcount\MIN   \MIN=\the\time  \advance\MIN by -\HOUR  \divide\HOUR by 60
   \ifnum  \MIN>9  \def\printTIME{{\it\the\HOUR\,:\,\the\MIN}}
   \else   \def\printTIME{{\it\the\HOUR\,:\,0\the\MIN}} 
   \fi 
   
   \def\LLab#1{\BP(0,0)\unitlength=1mm\put(-12,.5){\makebox(0,0)[cr]{\small #1
        \rlap{$_{_{\makeatletter\csname TRef#1\endcsname\makeatother}}$}}}\EP}
\fi

\begin{document}


{\hfill \archive   \vskip -2pt \hfill\preprint }
\vskip 15mm
\begin{center} 
{\huge\bf   \title }\vskip 10mm
Harald SKARKE\\[3mm] 
Mathematical Institute, University of Oxford,\\
24-29 St. Giles', Oxford OX1 3LB, England\\
{\tt skarke@maths.ox.ac.uk}

\vfill                  {\bf ABSTRACT } 
\end{center}    
\abstract

\vfill \noindent \preprint\\[5pt] \finished \vspace*{9mm}
\thispagestyle{empty} \newpage
\pagestyle{plain}

\newpage
\setcounter{page}{1}

\section{Introduction}

One of the cornerstones of the second string revolution is the
conjecture that there exist a non-perturbative duality between
heterotic string theory compactified on a four--torus and Type IIA
theory on a K3 surface \cite{HT}.
This conjecture is based on the equivalence of the moduli spaces,
the existence of soliton solutions in Type IIA that are believed to be
heterotic strings \cite{Sen1,HS}, relations to Type IIB on coinciding
5-branes \cite{BVS,OV} and arguments involving the algebras of BPS
states in both theories \cite{HM}. 
The most striking predicition of this duality is that Type IIA theory
should acquire non-perturbative gauge symmetries at special points in
the moduli space.
It was soon realized that the occurrence of gauge symmetry enhancement
should be connected with ADE type singularities in the
compactification geometry \cite{Wstd}.
As pointed out already in \cite{Wstd} there are problems with the
assumption that gauge symmetry enhancement might occur at the orbifold
point in the moduli space where the classical geometry becomes
singular, because at the orbifold point the conformal field theory
remains well behaved. 
Indeed it was shown in \cite{Aegs} that, at least in some cases, the
orbifold point and the point of gauge symmetry enhancement do not
coincide because of the discrepancy between classical geometry and
quantum geometry which leads to a non-vanishing $B$-field at the
orbifold point.

A non-perturbative gauge symmetry of ADE type is assumed to arise in
the following manner.
The Cartan subgroup arises from the usual Kaluza--Klein modes, and the
roots of the Lie algebra come from D-brane states that become massless
at the point of gauge symmetry enhancement.
Apart from the aforementioned problem with the orbifold point the
picture of branes wrapping cycles of the classical geometry suffers
from the fact that the classical geometry of the resolution of the
singularity contains only the cycles corresponding to the simple
roots.
This raises the question of why certain linear combinations of large
volume D-brane states (those whose coefficients are the ones that
express positive roots in terms of simple roots) should be considered
to be independent states and others not.

This is the problem that we want to address in this paper.
We examine the moduli space of Type IIA compactified on a
resolution of $\IC^2/\IZ_n$ in detail, using local mirror symmetry and
results on central charges of D-branes.
In particular we give an explicit solution of the GKZ system
corresponding to this geometry.
This analysis leads to exact expressions for central charges and hence
quantum volumes of cycles.
We find that there is a point in the quantum K\"ahler moduli
space of the resolution where all $(n^2-n)/2$ branches of the
principal component of the discriminant locus meet. 
These branches and, moreover, certain D-brane states
becoming massless along them, are in one to one correspondence with the
complete set of positive roots of $A_{n-1}$.
As expected \cite{BVS} the monodromy group of these states when taken
around the various branches in the vicinity of the gauge enhancement
point is the Weyl group of $A_{n-1}$. 
Along the way we will also discuss certain issues related to McKay
correspondence. 
Roughly our results can be stated in the following way.
Classical holomorphic cycles (which are relevant in the large radius
limit) correspond to the simple roots of $A_{n-1}$ whereas the quantum
geometry features all positive roots.
While the orbifold point is connected with the representation theory
of $\IZ_n$ the point of enhanced gauge symmetry knows about $A_{n-1}$
and its Weyl group $S_n$.

The structure of this paper is as follows. 
In the next section we outline the general strategy for analyzing
D-brane central charges of toric Calabi--Yau spaces with the help of
local mirror symmetry and summarize certain expectations on these
models, in particular if they are of orbifold type.
In section three we perform this analysis for the case of the
resolution of $\IC^2/\IZ_n$.
We find that in order to solve the GKZ system it is useful to change
to coordinates different from the ones that are ususally used
(the transformation to these coordinates is given in an appendix).
We find the precise form of the matching between the classical D-brane
states at large volume and the quantum corrected central charges as
determined by the GKZ solutions.
After some remarks concerning conjectures related to McKay
correspondence we analyze the behaviour of the D-brane states in the
region where the various branches of the principal component of the
discriminant locus come together.
Finally a brief summary and discussion of these results is given.

\section{The quantum moduli space of a toric Calabi--Yau manifold}

In order to analyze the quantum moduli space and hence the spectrum of
D-brane states and central charges of a toric Calabi--Yau manifold $X$
it is useful to follow the steps outlined here.
This is a summary of various useful results and conjectures on local
mirror symmetry, D-brane central charges, monodromy and McKay
correspondence; for a more detailed presentation in a similar context
see \cite{dnc}.
\begin{enumerate}
\item 
Analyze the geometry of $X$; in particular, find the Mori cone (the
cone of effective curve classes) and its dual, the K\"ahler cone.
There exists an efficient toric method for doing this.
Determine the toric data of the local mirror $\tilde X$ from
those of $X$.
This amounts to using the monomial-divisor mirror map to assign a
monomial $M_i$ to every toric divisor $D_i$ of $X$.
Then $\tilde X$ is the hypersurface determined by a general polynomial
$\sum a_iM_i$ that is a sum of these monomials.
The linear equivalence relations among the toric divisors of $X$, as
encoded in the secondary fan of $X$, determine multiplicative
relations among the coefficients $a_i$, ensuring that the
complex structure moduli space of $\tilde X$ again allows a toric
description in terms of the variety determined by the secondary fan.
\item The quantum moduli space of complexified K\"ahler classes on $X$
is identified with the complex structure moduli space of $\tilde X$.
As shown in \cite{CKYZ} the quantum volumes of compact cycles in $X$
have to fulfill the GKZ differential system corresponding to $\tilde
X$.
More generally the central charges of D-branes on $X$ wrapping even
cycles must have this property.
\item In order to match the GKZ solutions with the data of the
classical geometry of $X$, consider the large volume limit on $X$
which corresponds to the large complex structure limit on $\tilde X$.
There is a simple procedure for determining the large complex
structure coordinates $z_i$ that vanish in this limit in terms of the
homogeneous moduli space coordinates $a_i$ by using toric data
associated with the Mori cone of $X$.
Depending on the level of sophistication, a D-brane may be described
as a holomorphic cycle (possibly with a vector bundle on it), a
K-theory class, or an object in the derived category of coherent
sheaves.
Independently of the chosen description it is always possible to
assign a Chern character to a D-brane, which is all we need in the
following. 
In the large volume limit the central charge of a D-brane $S$ is
determined by \cite{DR,DD}
\beq 
Z^{\rm lv}(t_i;S)=
-\int_Xe^{-\sum t_i T_i}{\rm ch}(S)\sqrt{{\rm Td}(X)},
\eeql{zlv}
where $\sum t_i T_i$ is the expansion of the complexified K\"ahler
class $B+iJ$ in terms of the generators $T_i$ of the K\"ahler cone. 
In particular we may take $S$ to be the push-forward (to $X$) of the
structure sheaf of some cycle in $X$.
Choosing such cycles to be the curves $C_i$ dual to the complexified
K\"ahler classes $T_i$ leads to 
\beq t_i-1={\ln z_i\over 2\pi i}+O(z).\eeql{matching}
Given a basis of GKZ solutions in terms of the $z_i$ (which behave like
polynomials in the $\ln z_i$ near $z=0$) this affords a matching between
geometrical objects (or D-branes) and the corresponding central charges.
\item If the GKZ solutions are not holomorphic along some codimension
one locus in the moduli space, they may undergo monodromy.
Along the large complex structure divisors $z_i=0$ classical reasoning
applies and we have $t_i\to t_i+1$, corresponding to tensoring $S$
with $\co(T_i)$.
More importantly, there is a codimension one locus determined by a
polynomial equation in the $a_i$ called the principal component of the
discriminant locus along which $\tilde X$ develops singularities.
There is a natural pairing 
\beq \<R,S\>=\int_X {\rm ch}(R^*)~{\rm ch}(S)~{\rm Td}(X),\eeql{pairing}
between K-theory classes $R$ and $S$ on a Calabi--Yau manifold $X$; if
$X$ is non-compact at least one of these classes should be compactly
supported. 
It is believed that monodromy around some branch of the principal
component is determined by a Fourier-Mukai type transformation 
\beq {\cal F}\to {\cal F}-\<S, {\cal F}\>S, \eeql{konmon}
where $S$ is a suitable object that becomes massless along this branch.
In particular, for compact $X$ it is believed that $S=\co_X$, the
structure sheaf of $X$.
\end{enumerate}

If $X$ is a resolution of a space of the form $\IC^d/G$ with $d=2$
or $d=3$ and $G$ a finite abelian group (and also in several other
cases), we may apply results or conjectures related to McKay
correspondence to say more about such models.
In such a case there exists a distinguished basis $\{R_i\}$ of
line bundles on $X$ whose sections transform like the characters of
$G$ under the action of $G$ \cite{GSV,IN}.
The pairing $\<~~,~~\>$ of (\ref{pairing}) leads to a natural basis
$\{S_i\}$ of compactly supported K-theory classes fulfilling
$\<R_i,S_j\>=\d_{ij}$. 

Another feature of such models is the existence of a distinguished
point in the moduli space called the orbifold point, corresponding to
the locus where all exceptional divisors are blown down in the {\it
classical} geometry.
It is mirror to the point where all $a_i$ corresponding to the
exceptional divisors under the monomial-divisor mirror map vanish.
The moduli space is singular at the orbifold point, and in its
vicinity it is possible to define a non-simply connected path that
leads to `orbifold monodromy'.
For the relation between the $S_i$ and the monodromies the following 
conjectures exist:
\begin{itemize}
\item At the orbifold point the $S_i$ correspond to the `fractional
branes' \cite{DD}; thus they should form a representation of the group
of quantum symmetries under orbifold monodromy. 
\item At some branch of the principal component of the discriminant
locus $S_0$ becomes massless and generates the monodromy around this
branch.
Often the same is true for the other $S_i$ and different branches \cite{dnc}.
\end{itemize}

\section{D-branes on the resolution of $\IC^2/\IZ_n$}

In this section we want to apply the general strategy outlined above
to the case where $X$ is the resolution of $\IC^2/\IZ_n$.
Despite the somewhat unusual nature of the local mirror geometry which
is zero dimensional in this case, the machinery of using the GKZ
system to calculate quantum volumes and central charges should work.

\subsection{The geometry of $X$ and $\tilde X$}

$\IC^2/\IZ_n$ can be resolved by the introduction of a set
$\{C_1,\ldots,C_{n-1}\}$ of exceptional curves.
The classes of these curves generate the Mori cone of the resolution
$X$. 
Intersection numbers between the $C_i$ are given by 
$C_i\cdot C_j=-A_{ij}$ where $A$ is the Cartan matrix of $SU(n)$.

The mirror geometry of this model was mentioned already in \cite{KMP}.
It can easily be
determined by toric methods (see \cite{CKYZ} for a general discussion
of local mirror symmetry and \cite{dnc} for a derivation in the
present context).
The polynomial that determines the local mirror is given by
\beq P=a_0y_1^n+a_1y_1^{n-1}y_2+\cdots +a_ny_2^n. \eeq
The $a_i$ correspond under the monomial-divisor mirror map to the
curves $C_i$ where $C_0$ and $C_n$ can be thought of as the vanishing
loci of the coordinates of the orbifolded $\IC^2$; therefore we will
assume $a_0\ne 0$ and $a_n\ne 0$.
The local mirror $\tilde X$ is the vanishing locus of $P$, i.e. it is
just a collection of $n$ points in the $\IP^1$ with homogeneous
coordinates $(y_1:y_2)$. 
It is the analogue of the Seiberg-Witten curve as constructed from local
mirror symmetry in \cite{KKV}.
A `singularity' of $\tilde X$ occurs whenever two or more of these points
coincide.
We will be interested in the complex structure moduli space of $\tilde
X$ which clearly does not change under redefinitions $y_1\to\l_1y_1$
and $y_2\to\l_2y_2$, implying that we should identify the moduli space
coordinates $\{a_i\}$ with $\{\l_1^{n-i}\l_2^i~a_i\}$ (more precisely,
these transformations are the analogues of what we would get for higher
dimensional theories where the concept of complex structure makes more
sense).

\subsection{The GKZ system and its solutions}

The GKZ operators corresponding to the Mori cone generators are
\beq\partial_{a_0}\partial_{a_2}-\partial_{a_1}^2,~~~
\partial_{a_1}\partial_{a_3}-\partial_{a_2}^2,~~\ldots,~~
\partial_{a_{n-2}}\partial_{a_n}-\partial_{a_{n-1}}^2.\eeq
If we also consider GKZ operators corresponding to elements of the
Mori cone that are not generators we find that any equation of the
type
\beq(\partial_{a_i}\partial_{a_j}-\partial_{a_k}\partial_{a_l})\Pi=0~~~
\hbox{ with }~~~i+j=k+l  \eeql{gkz}
belongs to the GKZ system.
As far as I am aware this system has been solved only for the case of
$n=2$ \cite{AGM}.
The standard approach towards working with it would be to
transform it to the large complex structure coordinates $z_i$.
We find, however, that it is more useful to consider it in terms of 
the zeroes of $P$.
Defining $b_i=a_i/a_0$ and $x=y_1/y_2$ we get
\beq x^n+b_1x^{n-1}+\cdots +b_n=(x-x_1)(x-x_2)\cdots(x-x_n)=0, \eeql{xb}
with
\beq b_1=-x_1-\ldots -x_n,~~~b_2=x_1x_2+x_1x_3+\ldots
+x_{n-1}x_n,~~~\ldots, ~~~b_n=(-1)^nx_1x_2\cdots x_n.\eeq
We have not used the complete
freedom of redefinitions, the remaining freedom implying that we
should identify $(x_1,\ldots, x_n)$ with $(\l x_1,\ldots,\l x_n)$ for
any $\l\ne 0$.
In addition we should identify $(x_1,\ldots, x_n)$ with any
permuted version of itself.
The non-vanishing of $a_0$ and $a_n$ implies $x_i\ne 0$ and 
$x_i\ne \infty$, i.e. the moduli space is $(\IC^*)^n$ divided by
$\IC^*$ and $S_n$.

In order to express the GKZ system in terms of the $x_i$ we first note
that the identification of  $(x_1,\ldots, x_n)$ with 
$(\l x_1,\ldots,\l x_n)$ implies that any solution $\P$ of the GKZ 
system should be homogeneous in the $x_i$, i.e. it should fulfill
\beq x_1{\6\P\0\6 x_1}+\cdots+x_n{\6\P\0\6 x_n}=0.   \eeql{homog}
In the appendix we show that eqs. (\ref{gkz}) hold if and only if
\beq {\6^2\P\0\6 x_i \6 x_j}=0~~~\hbox{ for any }i\ne j.\eeql{dxidxj}
This set of equations must have a basis of solutions
$f_1(x_1),\ldots,f_n(x_n)$ each of which depends only on one of the
$x_i$.
By (\ref{homog}) it is then clear that $x_idf_i/dx_i=$const., i.e.
$f_i=\a+\b \ln x_i$ with $\a$, $\b$ constants.
The complete set of solutions of the combined system
(\ref{homog}, \ref{dxidxj}) is thus generated (redundantly) by
\beq \P_0=1,~~~\P_{ij}={1\0 2\p i}\ln\({x_i\0 x_j}\),~~i\ne j.\eeql{gkzsol}

\subsection{D-brane states and their central charges}

We now want to find the precise form of the mirror map by considering
the large volume limit of $X$ and the large complex structure limit of
$\tilde X$.
The large complex structure coordinates on $\tilde X$ are given by 
$z_i=a_{i-1}a_{i+1}/a_i^2$.
Let us assume that there exists a strong hierarchy between the
$x_i$ in the sense that $|x_{i+1}/x_i|<\d$ for all $x_i$ with some
small number $\d$.
Then we have $b_i=(-1)^ix_1x_2\cdots x_i(1+\co(\d))$ and
\beq z_i={a_{i-1}a_{i+1}\0 a_i^2}={b_{i-1}b_{i+1}\0 b_i^2}=
{x_{i+1}\0 x_i}(1+\co(\d)).  \eeq
Thus it is clear that the limit $|x_{i+1}/x_i|\to 0$ for all $i$ is
the large volume limit; in this limit 
\beq \ln z_i=\ln\({x_{i+1}\0 x_i}\)+\co(z).  \eeql{lnzlnx}

As the Mori cone is dual to the K\"ahler cone w.r.t. intersection
pairing, a basis $\{[C_i^\vee]\}$ of divisor classes dual to $\{C_i\}$
must form a basis for the K\"ahler cone.
We can now use (\ref{zlv}) (with $T_i=C_i^\vee$), (\ref{matching}) and
(\ref{lnzlnx}) to determine the exact central charges of D-brane states.
In terms of a basis $\{p,C_1,\ldots,C_{n-1}\}$ of compact cycles where
$p$ denotes the class of a point, we find 
\bea {\rm ch}(S)=p:&&Z^{\rm lv}(t_i;S)=-1~~\then~~Z(S)=-1,\\
{\rm ch}(S)=C_i:&&Z^{\rm lv}(t_i;S)=t_i~~\then~~
Z(S)={1\0 2\p i}\ln\({x_{i+1}\0 x_i}\)+1.\eea
If we denote by $\co_p$, $\co_{C_i}$ the structure sheaves of the
corresponding cycles as well as the sheaves on $X$ obtained by pushing
forward these structure sheaves, we find with the help of the
Grothendieck-Riemann-Roch formula 
\beq {\rm ch}(\co_{C_i})=p+C_i,~~~ {\rm ch}(\co_p)=p  \eeq
and therefore 
\beq Z(\co_{C_i})={1\0 2\p i}\ln\({x_{i+1}\0 x_i}\),~~~
Z(\co_p)=-1.\eeq

\subsection{McKay correspondence and orbifold monodromy}

As a brief interlude we will now check that the $S_i$
get permuted cyclically by orbifold monodromy $a_j\to e^{2\pi i~j/n}
a_j$ corresponding to $x_i\to e^{2\pi i/n}x_i$. 
The line bundles $R_i$ that form the McKay basis of $K(X)$ are given
by $R_0=\co_X$ and $R_i=\co_X(C_i^\vee)$ for $i\ge 1$ \cite{GSV}.  
By applying (\ref{pairing}) we find
\beq {\rm ch}(S_0)=p+\sum_{i=1}^{n-1}C_i,~~~
{\rm ch}(S_i)=- C_i~~\hbox{for}~~i>0 \eeq
leading to
\beq Z(S_0)=n-2+\sum_{i=1}^{n-1}{1\0 2\p i}\ln\({x_{i+1}\0 x_i}\),~~
Z(S_i)=-{1\0 2\p i}\ln\({x_{i+1}\0 x_i}\)-1~~\hbox{for}~~i>0.\eeq
Clearly $\sum_{i=0}^n Z(S_i)=-1$ everywhere in the moduli space.
The orbifold point is the fixed point of the orbifold monodromy
determined by $a_i=0$ for $1\le i\le n-1$,
i.e. the set of $x_i$ is just the set of $n$-th unit roots (up to a
common multiplicative constant).
Provided we choose a suitable path from the large complex structure
region to the orbifold point, we can get 
$x_{i+1}/x_i=e^{-2\pi i(n-1)/n}$ and thus $Z(S_i)=-1/n$ for each $i$.
With this choice $Z(S_0)=-(1/2\p i)\ln(x_{1}/x_n)-1$ and
orbifold monodromy indeed acts by shifting the $S_i$ cyclically
according to the defining representation of $\IZ_n$.

\subsection{Principal component monodromy}
The principal component of the discriminant locus in the moduli space
consists of the loci where $x_i=x_j$ for some $i\ne j$.
As the moduli space is the space of unordered n-tuples of $x_i$, i.e. 
the space of ordered n-tuples divided by the permutation group of the
$x_i$, monodromy around the locus $x_i=x_j$ results in exchanging
$x_i$ with $x_j$. 
In terms of the solutions (\ref{gkzsol}) of the GKZ system, this
monodromy acts by
\beq 
\P_{ij}\leftrightarrow -\P_{ij},~~~~~~\P_{ik}\leftrightarrow\P_{jk},~~~ 
\P_{ki}\leftrightarrow \P_{kj}~~~\hbox{ for $k$ neither $i$ nor $j$.}
\eeq
Let us compare this to the root system and Weyl group of $SU(n)$.
The roots $\a_{ij}$ of $SU(n)$ can be represented as $\a_{ij}=e_i-e_j$
where the $e_i$ are linearly independent vectors in an $n$ dimensional
space. 
In particular, the positive roots may be chosen to correspond to
$e_{i+1}-e_i$. 
A Weyl reflection through the plane
orthogonal to $\a_{ij}$ is determined by 
$v\to v-\<\a_{ij},v\>_{\rm C}~\a_{ij}$ 
with $\<~~,~~\>_{\rm C}$ the bilinear pairing that acts on two simple
roots by giving the corresponding element of the Cartan matrix.
This Weyl reflection is induced by an exchange of $e_i$ and $e_j$ and
the complete Weyl group corresponds to the permutation group of the
$e_i$.
This shows that the group of principal component monodromies of the
GKZ solutions is isomorphic to the Weyl group of $SU(n)$ with the
correspondence  ${1\0 2\pi i}\ln x_i\leftrightarrow e_i$.
Note that this fits well with the occurrence of braid group
relations \cite{ST} in the homological algebra approach to mirror
symmetry \cite{Kon}.

This is consistent with the Fourier-Mukai transformation (\ref{konmon}).
We have $Z(\co_{C_i})=0$ at
$x_i=x_{i+1}$, so we hope that $\co_{C_i}$ is the object generating
the monodromy around this locus (according to \cite{ST}, the
$\co_{C_i}$ are spherical and are thus the objects we expect to
generate monodromies of this type).
Using the pairing (\ref{pairing}) we find 
\beq \<\co_{C_i},\co_p\>=0~~\hbox{ and }~~
\<\co_{C_i},\co_{C_j}\>=-C_i\cdot C_j=A_{ij}\eeq
with $A$ being the Cartan matrix of $SU(n)$. 
By linearity we also get the right behavior at $x_i=x_j$ with $i>j$
where $Z(\co_{C_{i-1}}+\co_{C_{i-2}}+\cdots\co_{C_j})=0$.
The following table lists the correspondences between the $SU(n)$ Lie
algebra, the GKZ solutions and the D-brane states.

\ni\begin{tabular}{|l|l|l|} \hline
Lie algebra of $SU(n)$ & GKZ solution / $Z(S)$ & D-brane $S$ \\ \hline\hline
Simple root $\a_{i+1,i}=e_{i+1}-e_i$ & 
   $\P_{i+1,i}={1\0 2\pi i}\ln x_{i+1}-{1\0 2\pi i}\ln x_i$ & 
   $\co_{C_i}$ \\ \hline
Positive root $\a_{ij}=e_i-e_j$ & 
   $\P_{ij}={1\0 2\pi i}\ln x_i-{1\0 2\pi i}\ln x_j$ & \\
$~~~~~~=\a_{i,i-1}+\ldots +\a_{j-1,j}$ & 
   $~~~~~=\P_{i,i-1}+\ldots +\P_{j-1,j}$ & 
   $\co_{C_{i-1}}+\ldots +\co_{C_j}$  \\ \hline
Weyl reflection $e_i\leftrightarrow e_j$ & 
  $\P_{ij}\leftrightarrow -\P_{ij},~\P_{ik}\leftrightarrow\P_{jk},~
     \P_{ki}\leftrightarrow \P_{kj}$ & \\
$~~~~~~v\to v-\<\a_{ij},v\>\a_{ij}$ & & 
  ${\cal F}\to {\cal F}-\<S, {\cal F}\>S$  \\ \hline
Weyl group: Permutations of $e_i$ & Permutations of $x_i$ &  \\ \hline
\end{tabular}

\section{Summary and discussion}
We have seen that it is possible to describe the moduli space of a
resolution of $\IC^2/\IZ_n$ in terms of an unordered set
$\{x_1,\ldots,x_n\}$ of points in $\IC^*$, up to a common non-vanishing
factor, and to solve the GKZ system explicitly in terms of the $x_i$.
There are two very special points in this moduli space. 
One of them is the orbifold point where the $x_i$ are just $n$'th
roots of unity.
The other one is the point of maximal gauge symmetry enhancement where
all of the $x_i$ come together.
Our explicit analysis confirms the picture of the orbifold point as
the locus in the moduli space where `fractional branes' with charge
$1/n$ times the charge of a D0-brane occur and where McKay
correspondence (i.e., connections between the representation theory of
the finite orbifolding group and the geometry of the resolution) plays
a role. 
At the point of enhanced gauge symmetry we find that all $n(n-1)/2$
branches of the discriminant locus meet.
At each of these branches a specific quantum cycle becomes massless
(leading to an $SU(2)$ group), and there is a one to one
correspondence of K-theory classes with vanishing central charge at
these cycles and combinations of large volume cyles with coefficients
that are just the ones that determine the positive roots of $SU(n)$ in
terms of the simple ones.
This provides a simple answer to the question of which states of
wrapping branes should be considered to be independent:
there is one state for every (anti-)brane wrapping a cycle in the
quantum geometry.

Our picture of the moduli space near the point of gauge enhancement 
(with all $x_i$ coming together) is very suggestive of a dual type IIB
description where the gauge group would be enhanced by strings
becoming massless at the points in moduli space where branes come
together. 
We note, however, that this picture is valid only near the gauge
enhancement point and not globally. 
For example, it could not be used to describe the orbifold point.

The very explicit nature of the present description of the quantum moduli
space and the central charges should make it a useful testing ground
for many ideas related to finding a precise mathematical formulation
of D-brane states.
We have seen already how the braid group structures that are expected
to arise in the context of derived categories manifest themselves as
monodromies. 
Having a closed formula for the periods should also make it easy to
discuss the $\Pi$-stability of \cite{DFR,Dou}.

\section*{Appendix}
In this appendix we show how to transform the GKZ system as formulated
in (\ref{gkz}), involving the $a_i$, into a system involving the
$x_i$.
In order to express $\6_{b_j}$ in terms of $\6_{x_i}$ we have to
invert the matrix $C$ with $C_{ij}=\6 b_j/\6x_i$.
Differentiating (\ref{xb}) by $x_i$ implies
\beq \sum_{j=1}^n x^{n-j} {\6 b_j\0 \6x_i}=-\prod_{k\ne i} (x-x_k). 
\eeql{diff}
Introducing the matrix $D$ with elements $D_{jk}=x_k^{n-j}$ and the
notation 
\beq \p_i:=\prod_{k\ne i}(x_i-x_k)\eeq
we get $C\cdot D=-~{\rm diag}(\p_1,\ldots,\p_n)$ and thus 
$(C^{-1})_{ji}=-x_i^{n-j}/\p_i$ and 
$\6_{b_j}=-\sum_i (x_i^{n-j}/ \p_i)\6_{x_i}$.
In terms of the original moduli space coordinates $a_i$ this implies
\bea a_0 \6_{a_0}&=&
-\sum_{j=1}^n b_j\6_{b_j}=
\sum_{j=1}^n b_j \sum_i {x_i^{n-j}\0 \p_i}\6_{x_i}=
\sum_i{\sum_{j=1}^n b_j x_i^{n-j}\0 \p_i}\6_{x_i}=
-\sum_i{x_i^n\0 \p_i}\6_{x_i},\\
a_0\6_{a_j} &=& \6_{b_j}=-\sum_i {x_i^{n-j}\0 \p_i}\6_{x_i}
~~\hbox{ for }~~j\ge 1. \eea
Now (\ref{gkz}) can be expressed as
\beq 
\sum_{i=1}^n{x_i^p\0 \p_i}\6_{x_i}
\(\sum_{j=1}^n{x_j^q\0 \p_j} \6_{x_j}\P\)=
\sum_{i=1}^n{x_i^r\0 \p_i}\6_{x_i}
\(\sum_{j=1}^n{x_j^s\0 \p_j} \6_{x_j}\P\)
\eeql{DPiDPi}
for all $p,r\in\{0,1,\ldots,n-1\}$, $q,s\in\{0,1,\ldots,n\}$ with
$p+q=r+s$. 
Clearly each side of eq. (\ref{DPiDPi}) can be written as 
$\sum_jA_j\6_{x_j}\P + \sum_{i,j}B_{ij}\6_{x_i}\6_{x_j}\P$.
We now want to show that the $A_j$-terms on both sides are the same,
i.e. cancel one another.
We find 
\beq A_j=\sum_{i=1}^n{x_i^p\0 \p_i}\6_{x_i} \({x_j^q\0 \p_j}\)=
{x_j^q\0 \p_j}
\({x_j^p\0 \p_j}\({q\0 x_j}-\sum_{k\ne j}{1\0 x_j-x_k}\)
+\sum_{i\ne j}{x_i^p\0 \p_i}{1\0 x_j-x_i}\).
\eeq
The last expression can be transformed as
\bea \sum_{i\ne j}{x_i^p\0\p_i(x_j-x_i)}&=&
\sum_{i\ne j}\(-{x_i^{p-1}\0 \p_i}+
{x_jx_i^{p-1}\0 \p_i(x_j-x_i)}\)=
{x_j^{p-1}\0 \p_j}+x_j\sum_{i\ne j}{x_i^{p-1}\0 \p_i(x_j-x_i)}=
\cdots =\nn\\
&&p{x_j^{p-1}\0 \p_j}+x_j^p\sum_{i\ne j}{1\0 \p_i(x_j-x_i)},
\eea
where we have made use of $\sum_{i=1}^n x_i^k/\p_i=0$ for 
$k<n-1$ (this can be seen by noting that it would have to be a
rational function of negative degree but none of the possible poles
actually is a pole).
Thus we arrive at
\beq A_j={x_j^{p+q}\0 \p_j^2}
\({p+q\0 x_j}+\sum_{i\ne j}{\p_j-\p_i\0 \p_i(x_j-x_i)}\),
\eeq
which depends on $p$ and $q$ only via $p+q$, ensuring that
because of $p+q=r+s$ the corresponding terms drop out from
(\ref{DPiDPi}) which becomes
\beq 
\sum_{i=1}^n\sum_{j=1}^n{x_i^px_j^q-x_i^rx_j^s\0 \p_i\p_j}
{\6^2\P\0\6x_i\6x_j}=0. \eeq
Choosing $r=p+1$, $q=s+1$ and $s>p$ (equations corresponding to different
values can be obtained by linear combinations of these) and
reshuffling the summation, we get
\bea 0&=& \sum_{1\le i<j\le n}
{x_i^px_j^{s+1}-x_i^{p+1}x_j^s+x_j^px_i^{s+1}-x_j^{p+1}x_i^s\0 \p_i\p_j}
{\6^2\P\0\6x_i\6x_j}=\nn\\
&&\sum_{1\le i<j\le n}
{(x_i-x_j)^2\0 \p_i\p_j}
(x_i^{s-1}x_j^p+x_i^{s-2}x_j^{p+1}+\ldots +x_i^px_j^{s-1})
{\6^2\P\0\6x_i\6x_j}=:\nn\\
&&\sum_{1\le i<j\le n} M_{ps,ij}{\6^2\P\0\6x_i\6x_j}
~~\hbox{ for all }~~0\le p<s\le n-1. \label{MPI}\eea
We now want to show that the ${n\choose 2}\times {n\choose 2}$ matrix
$M$ acting on the ${n\choose 2}$ vector $\6^2\P/(\6x_i\6x_j)$ is
regular for generic values of the $x_i$.
This is true iff the matrix $\tilde M$ with entries 
$x_i^{s-1}x_j^p+\ldots +x_i^px_j^{s-1}$ is regular (the other factor
can be pulled out in calculating the determinant of $M$).
The non-vanishing of the determinant of $\tilde M$ can be shown
inductively.
It is true for $n=2$ and it is not hard to see that
\bea \det\tilde M^{(n)}(x_1=0,x_2,\ldots,x_n)=\det D(x_2,\ldots,x_n)~
(x_2\cdots x_n)^{n-2}~\det\tilde M^{(n-1)}(x_2,\ldots,x_n)\ne 0,\nn\eea
where $D$ is the regular matrix we encountered after eq. (\ref{diff}).
Thus $M$ is regular, and by (\ref{MPI}) we have completed the
proof that (\ref{gkz}) is equivalent to (\ref{dxidxj}).


\small

\begin{thebibliography}{11}

\def\I#1{{\it #1}}      \addtolength{\itemsep}{-4.5pt}  \small \vspace{-3mm}

\ifundefined{draftmode} \def\.#1 #2\>{\bibitem{#1}#2}           
\else                   \def\.#1 #2\>{\bibitem{#1}\LLab{#1}#2}  \fi

\.HT    C. Hull and P. Townsend, \I{Unity of Superstring Dualities,}
        \npb 438 (1995) 109, hep-th/9410167. \>

\.Sen1  A. Sen, \I{String String Duality Conjecture in Six Dimensions
        and Charged Solitonic Strings,} \npb 450 (1995) 103, hep-th/9504027.\> 

\.HS    J.A. Harvey, A. Strominger, \I{The Heterotic String is a 
        Soliton,}
        \npb 449 (1995) 535, Erratum-ibid. B458 (1996) 456, hep-th/9504047.  \>

\.BVS   M. Bershadsky, C. Vafa and V. Sadov, \I{D-Strings on D-Manifolds,}
        \npb 463 (1996) 398,   hep-th/9510225. \>

\.OV    H. Ooguri, C. Vafa, \I{Two-Dimensional Black Hole and
	Singularities of CY Manifolds,} hep-th/9511164, 
	\npb 463 (1996) 55.   \>

\.HM    J. Harvey, G. Moore, \I{On the algebra of BPS states,} \cmp
        197 (1998) 489, hep-th/9609017. \>

\.Wstd  E. Witten, \I{String Theory Dynamics in Various Dimensions,}
        \npb 443 (1995) 85, hep-th/9503124. \>

\.Aegs  P.S. Aspinwall, \I{Enhanced Gauge Symmetries and K3 Surfaces,}
        \plb 357 (1995) 329, hep-th/9507012.  \>

\.dnc   X. de la Ossa, B. Florea and H. Skarke, \I{D-Branes on
	Noncompact Calabi-Yau Manifolds: K-Theory and Monodromy,} 
	hep-th/0104254.    \>

\.CKYZ  T.-M. Chiang, A. Klemm, S.-T. Yau, and E. Zaslow, 
	\I{Local Mirror Symmetry: Calculations and Interpretations,} 
	\atmp 3 (1999), 495, hep-th/9903053.   \>

\.DR    D.-E. Diaconescu and C. R\"omelsberger, \I{D-branes and 
        bundles on elliptic fibrations,} {\it Nucl. Phys.} {\bf B574} (2000) 
        245, hep-th/9910172.   \>

\.DD    D.-E. Diaconescu and M. R. Douglas, \I{D-branes on stringy
	Calabi-Yau manifolds,} hep-th/0006224.   \>

\.GSV   G. Gonzales-Sprinberg and J.-L. Verdier, \I{Construction
	geometrique de la correspondance de McKay,} Ann. sci. ENS 
	{\bf 16} (1983), 409.   \>

\.IN    Y. Ito and H. Nakajima, \I{McKay correspondence and Hilbert
	schemes in dimension three}, math.AG/9803120.  \>

\.KMP   S. Katz, D.R. Morrison and M.R. Plesser, \I{Enhanced Gauge Symmetry 
        in Type II String Theory,} \npb 477 (1996) 105, hep-th/9601108. \>

\.KKV   S. Katz, A. Klemm, C. Vafa, \I{Geometric Engineering of Quantum 
        Field Theories,} \npb 497 (1997) 173, hep-th/9609239.\>

\.AGM   P. S. Aspinwall, B. R. Greene, and D. R. Morrison,
	\I{Measuring small distances in $N=2$ sigma models,} 
	{\it Nucl. Phys.} {\bf B420} (1994) 184, hep-th/9311042.  \>

\.ST    P. Seidel and R.P. Thomas, \I{Braid group actions on
	derived categories of coherent sheaves,} math.AG/0001043.    \>  

\.Kon   M. Kontsevich, \I{Homological Algebra of Mirror Symmetry},
	alg-geom/9411018. \>

\.DFR   M. R. Douglas, B. Fiol and C. R\"omelsberger, \I{Stability and
	BPS branes,} hep-th/0002037. \>

\.Dou   M. R. Douglas, \I{D-branes, Categories and N=1 Supersymmetry,}
	hep-th/0011017.   \>

\end{thebibliography}
\bye